# An investigation of the false discovery rate and the misinterpretation of *P* values

## David Colquhoun

Neuroscience, Physiology & Pharmacology. University College London, Gower Street, WC1 6BT

Email: d.colquhoun@ucl.ac.uk

## Abstract

If you use *P* = 0.05 to suggest that you have made a discovery, you'll be wrong at least 30% of the time.  If, as is often the case, experiments are under-powered, you'll be wrong most of the time.  This conclusion is demonstrated from several points of view. First, tree diagrams which show the close analogy with the screening test problem. Similar conclusions are drawn by repeated simulations of *t* tests.  These mimic what's done in real life, which makes the results more persuasive  The simulation method is used also to evaluate the extent to which effect sizes are over-estimated, especially in under-powered experiments. A script is supplied to allow the reader to do simulations themselves, with numbers appropriate for their own work.  It is concluded that if you wish to keep your false discovery rate below 5%, you need to use a 3-sigma rule, or to insist on *P* ≤ 0.001.  And *never* use the word "significant".

*'. • • before anything was known of Lydgate's skill, the judgements on it had naturally been divided, depending on a sense of likelihood, situated perhaps in the pit of the stomach or in the pineal gland, and differing in its verdicts, but not less valuable as a guide in the total deficit of evidence. '*George Eliot (Middlemarch, Chap. 45)

*"The standard approach in teaching, of stressing the formal definition of a p-value while warning against its misinterpretation, has simply been an abysmal failure"* Sellke *et al.* (2001) `

## Introduction

There has been something of a crisis in science. It has become apparent that an alarming number of published results can't be reproduced by other people. That's what caused John Ioannidis to write his now famous paper, <u>Why Most Published Research Findings Are False</u> (Ioannidis, 2005). That sounds very strong. But in some areas of science it is probably right. One contribution to this sad state of affairs must be the almost universal failure of biomedical papers to appreciate what governs the false discovery rate.

In 1971, my point of view was that

*"the function of significance tests is to prevent you from making a fool of yourself, and not to make unpublishable results publishable" (Colquhoun, 1971).*

(Now, of course, one appreciates better the importance of publishing all results, whether negative or positive.)

You make a fool of yourself if you declare that you have discovered something, when all you are observing is random chance. From this point of view, what matters is the probability that, when you find that a result is "statistically significant", there is actually a real effect. If you find a "significant" result when there is nothing but chance at play, your result is a false positive, and the chance of getting a false positive is often alarmingly high. This probability will be called *false discovery rate* in this paper. It's also often called the *error rate*.

You can also make a fool of yourself if you fail to detect a real effect, though that is less bad for your reputation.

The false discovery rate is the complement of the *positive predictive value* (PPV) which is the probability that, when you get a "significant" result there is actually a real effect. So, for example, if the false discovery rate is 70%, the PPV is 30%. The false discovery rate is a more self-explanatory term so it will be preferred here.

If you are foolish enough to define "statistically significant" as anything less than $P = 0.05$ then, according to one argument (Sellke *et al.*, 2001) ), you have a 29% chance (at least) of making a fool of yourself. Who would take a risk like that? Judging by the medical literature, most people would. No wonder there is a problem.

The problems can be explained easily without using any equations, so equations are confined to the appendix, for those who appreciate their beauty. Before talking about significance tests, it will be helpful to reiterate the problem of false discovery rate in screening tests. Although this has had much publicity recently, it is not widely appreciated that very similar arguments lead to the conclusion that tests of significance are misinterpreted in most of the biomedical literature.

## The screening problem

The argument about false positives in significance testing is closely related to the argument about false positives in diagnostic tests. The latter may be a bit more familiar, so let's deal with it first.

Imagine a test for which 95 percent of people without the condition will be correctly diagnosed as not having it (specificity = 0.95). That sounds pretty good. Imagine too, that the test is such that four out of five people with the condition will be detected by this test (sensitivity = 0.8). This sounds like a reasonably good test. These numbers are close to those that apply for a rapid screening test for Alzheimer's disease, proposed by Scharre *et al.* (2014). Or, rather, it's a test for mild cognitive impairment, MCI, which may or may not develop into dementia. It's been proposed that everyone should get dementia screening. Is this sensible? One could argue that even if the test were perfect, screening would be undesirable because there are no useful treatments. But it turns out that the test is almost useless anyway. We need to know one more thing to find the false discovery rate, namely the prevalence of MCI in the population to be screened. For the whole population this is just over 1% (or about 5% for people over 60). Now we can construct the diagram in Fig 1.

If we screen 10,000 people, 100 (1%) will have MCI, and 9,900 (99%) will not. Of the 9,900 people who At this point I should clarify that this paper is not about multiple comparisons. It's well known that high false discovery rates occur when many outcomes of a single intervention are tested. This has been satirised as the "jelly bean" problem (http://xkcd.com/882/ ). Despite its notoriety it is still widely ignored. There exist several methods that compensate for the errors that occur as a result of making multiple comparisons. The best known is the Bonferroni correction, but arguably that method sets a criterion that is too harsh, and runs an excessive risk of not detecting true effects (it has low power). In contrast, the method of Benjamini and Hochberg (1995) is based on setting a limit on the false discovery rate, and this is generally preferable.

However this paper is not concerned with multiple comparisons. It deals only with the very simplest ideal case. We ask how to interpret a single $P$ value, the outcome of a test of significance. All of the assumptions of the test are true. The distributions of errors are precisely gaussian and randomisation of treatment allocations was done perfectly. The experiment has a single pre-defined outcome. The fact that, even in this ideal case, the false discovery rate can be alarmingly high means that there is a real problem for experimenters. Any real experiment can only be less perfect than the simulations discussed here, and the possibility of making a fool of yourself by claiming falsely to have made a discovery can only be even greater than we find in this paper.

The simplest way you can estimate your false discovery rate is very easy to follow.

The classical $P$ value does exactly what it says. But it is a statement about what would happen if there were no true effect. That can't tell you about your long-term probability of making a fool of yourself, simply because sometimes there really is an effect. In order to do the calculation we need to know a few more things.

First we need to know the probability that the test will give the right result when there is a real effect. This is called the power of the test. The power depends on the sample size, and on the size of the effect we hope to detect. When calculating sample sizes it is commonly set to 0.8, so when there is a real effect, we'll detect it (declare the result to be 'significant') in 80% of tests. (Clearly it would be better to have 99% but that would often mean using an unfeasibly large sample size.) The power of the significance test is the same thing as the sensitivity of a screening test.

There is one other thing that we need to specify in order to calculate the false discovery rate. We must take a guess at the fraction of tests that we do in which there is a real difference. Of course we can't usually know this, and the value will depend, among other things, on how good we are at guessing what experiments to do.

For example, if the tests were on a series of homeopathic "remedies", none would have a real difference because the treatment pills would be identical with the placebo pills. In this case every test that came out 'significant' would be a false positive so our false discovery rate would be 100%.

More realistically, imagine that we are testing a lot of candidate drugs, one at a time. It's sadly likely that many of them would not work, so let's imagine that 10% of them work and the rest are inactive.

We can work out the consequences of these numbers exactly as we did for the screening problem. This is shown in Fig 2.

Imagine that over a period of time you do 1000 tests. Of these, 100 (10%) will have real effects, and 900 (90%) will be cases where there is no real effect.

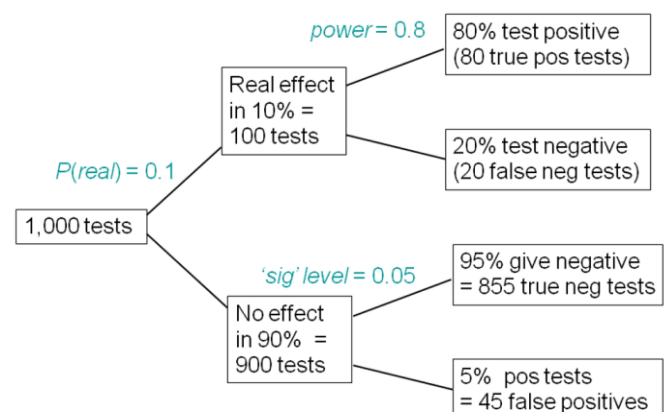

**Figure 2** Tree diagram to illustrate the false discovery rate in significance tests. This example considers 1000 tests, in which the prevalence of real effects is 10%. The lower limb shows that with the conventional significance level, $P = 0.05$, there will be 45 false positives. The upper limb shows that there will be 80 true positive tests. The false discovery rate is therefore 45/(45+80) = 36%, far bigger than 5%.

If we consider the 900 tests where there is no real effect (the null hypothesis is really true) then, according to the classical theory, 45 tests (5% of them) will be false positives (lower limb in Fig 2). So you might imagine that the false discovery rate is 5%. It isn't, because in order to work out the fraction of positive tests that give the wrong result, we need to know the total number of positive tests.

To find this we need to look also at the upper limb in Fig 2. In the 100 tests (10%) where there is a real effect (the null hypothesis is false), the effect will be correctly detected in 80 (80%) but 20 tests will fail to detect the effect (false negatives).

Thus the total number of positive tests is 80 + 45 = 125. Of these, 45 are false positives. So In the long run, your chance of making a fool of yourself by declaring a result to be real when it isn't will be 45/125 = 36%.

This false discovery rate is far bigger than 5%. It may go some way to explain why so many false positive tests corrupt the literature,

The approach that has just been described is sometimes described as Bayesian but notice that all the probabilities that are involved can be expressed in terms of long-run probabilities. It can be regarded as an exercise in conditional probabilities. The description 'Bayesian' is not wrong but it is not necessary.

### A few more complications

The argument outlined above is simple. It shows there is a problem, but doesn't provide all the answers. Once we go a bit further, we get into regions where statisticians disagree with each other.

*It is difficult to give a consensus of informed opinion because, although there is much informed opinion, there is rather little consensus. A personal view follows* (Colquhoun, 1971).

An easy way to test these problems is to simulate a series of tests to see what happens in the long run. This is easy to do and a script is supplied (Colquhoun, 2014b), in the R language. This makes it quick to simulate 100,000 *t* tests (that takes about 3.5 min on my laptop). It's convincing because it mimics real life.

Again we'll consider the problem of using a Student's *t* test to test whether there is a real difference between the means of two groups of observations. For each test two groups of simulated 'observations' are generated as random variables with specified means and standard deviations. The variables are normally distributed, so the assumptions of the *t* tests are exactly fulfilled.

When the true means of both groups are the same, the true mean difference between the means is zero. The distribution of the differences for 100,000 such tests is shown in Fig 3a. As expected, the average difference is close to zero. In this example each group had 16 observations with a standard deviation of 1 for both group, so the standard deviation for each mean (the standard error) is $1/\sqrt{16} = 0.25$, and the standard deviation of the difference between them is $\sqrt{(0.25^2 + 0.25^2)} = 0.354$. If the observations follow a normal (gaussian) distribution, and we use $P = 0.05$ as the threshold for 'significance', then we find that 5% of the tests will be 'significant', and all of these are false positives. This is all we need to know for the classical approach.

The distribution of the 100,000 *P* values generated is shown in Fig 3b: 5% of them (5000 values) are indeed below $P = 0.05$, but note that the distribution is flat (in statisticians' jargon, the distribution of *P* values under the null hypothesis is uniform). So there is the same number of *P* values between 0.55 and 0.6, and in every other interval of the same width. This means that *P* values are not at all reproducible: all values of *P* are equally likely.

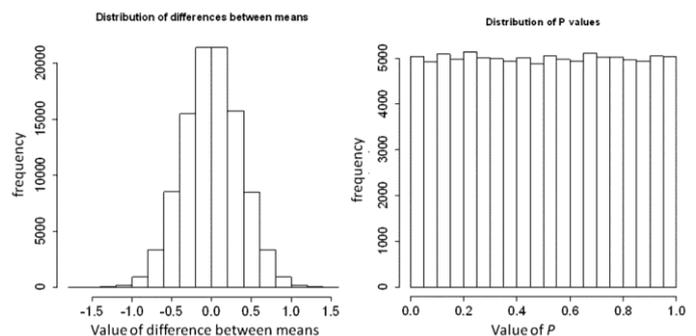

**Figure 3** Results of 100,000 simulated *t* tests, when the null hypothesis is true. The test looks at the difference between the means of two groups of observations which have identical true means, and a standard deviation of 1. Fig 3a shows the distribution os the 100,000 'observed' differences between means (it is centred on zero and has a standard deviation of 0.354). Fig 3b shows the distribution of the 100,000 *P* values. As expected. 5% of the tests give (false) positives ($P \le 0.05$), but the distribution is flat (uniform).

In order to complete the picture we need to consider also what happens when there is a real difference between the means. Suppose that the treatment group has a true mean that is greater than that of the control group by 1, so the true mean difference between groups is 1. The distributions of observations for control (blue) and treatment (red) groups are shown in Fig 4a.

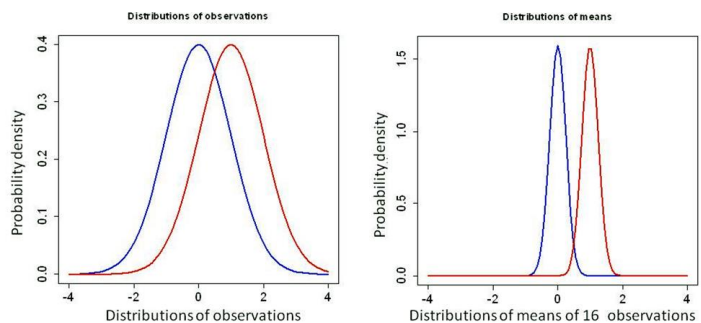

**Figure 4** The case where the null hypothesis is *not* true. Simulated *t* tests are based on samples from the postulated true distributions shown: blue = control group, red = treatment group. The observations are supposed to be normally distributed with means that differ by one standard deviation, as shown in in Fig 4a. The distributions of the means of 16 observations are shown in Fig 4b.

We've supposed that both groups have the same standard deviation, set to 1 in this example. (The exact numbers are not critical –the results will apply to any case where the true difference between means is one standard deviation.) The distributions for control and treatment groups show considerable overlap, but the means of 16 observations are less scattered. Their standard deviation is $1/\sqrt{16} = 0.25$.

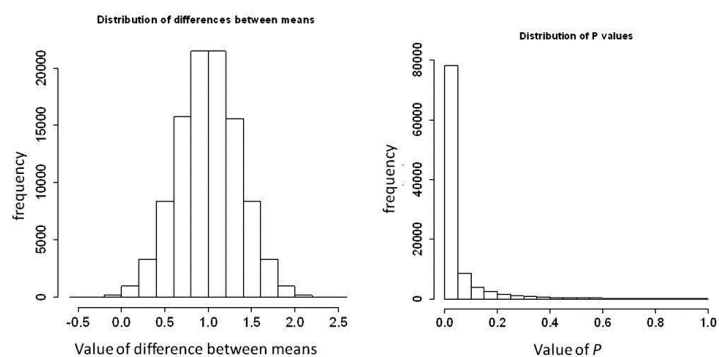

**Figure 5** Results of 100,000 simulated *t* tests in the case where the null hypothesis is *not* true, but as shown in Fig 4. Fig 5a shows the distribution of the 100,000 'observed' values for the differences between means of 16 observations test looks. It has a mean of 1, and a standard deviation of 0.354. Fig 5b shows the distribution of the 100,000 *P* values: 78% of them are equal to or less than 0.05 (as expected form the power of the tests).

The overlap is not huge. In fact the sample size of 16 was calculated to make the power of the test close to 0.8, so about 80% of real differences should be detected.

Fig 5a shows that the 'observed' differences are indeed centred around 1. Fig 5b shows that the number of *P* values that are equal to, or less than, 0.05, is 78%, as expected from the calculated power, 0.78 (there is a power calculator at http://www.stat.ubc.ca/~rollin/stats/ssize/n2.html ). In other words, 78% of tests give the correct result.

If we look at the average effect size for all those 'experiments' for which $P \le 0.05$, it is 1.14 rather than 1.0. The measured effect size is too big (see Fig 7), and this happens because experiments that have, by chance, a larger than average effect size are more likely to be found 'significant' than those that happen to have small effect size.

**The false discovery rate in the simulated *t* tests**

In order to work out the long-term chance of making a fool of yourself, we must postulate the fraction of experiments that we do where there is a true effect (e.g. a true difference between means in the simulations just described). This has already been done in a simple way in the tree diagram, in Figure 2. It is the equivalent of the prevalence in the screening example. Similar inferences can be made from the simulations. For the tree diagram we considered the case where 10% of all the experiments we do have real effects and 90% have no effects. We can take 90% of the simulations that had no true effect (Fig 3) and combine them with 10% of the simulations for which there was a true effect (Figs 4, 5). In 100,000 simulations, there are 5000 (5%) of false positives ($P \le$ 0.05) in Fig 3, and in Fig 5 there are 78,000 (78%) of (true) positives. Combining these gives 0.9 × 5,000 + 0.1 × 78,000 = 12,300 positive tests (*i.e.* those with $P \le$ 0.05), of which 5,000 are false positives. Therefore, if a positive test is observed, the probability that it is a false positive is (0.9 × 5,000) / 12,300 = 0.36.

Thus, you make a fool of yourself 36% of the time in this case, as inferred from the tree diagram in Fig 2. The false discovery rate is 36%, not 5%. The appendix shows how this number can be calculated from an equation, but there is no need for an equation to get the result. The R script (Colquhoun, 2014b) can be used to do simulations with your own numbers.

If we use a different postulate about the fraction of experiments in which there is no real effect, we get a different result.  For example, if we assume that there is a real effect in half the experiments we do, rather than in 10% of them, the example just used, only 6% of tests would be false positives, not much different from the *P* value of 0.05.  So in this particular case there seems to be no problem.  This doesn't get us off the hook, though, for three reasons. One reason is that there is no reason to think that half the tests we do in the long run will have genuine effects.  Another reason is connected with the rather subtle question of whether or not we should include *P* ≤ 0.05 in the calculation, when we have observed *P* = 0.05.  The third reason is that underpowered studies show a false discovery bigger than 0.05 even when there are 50% of experiments with genuine effects.  These questions will be considered next.

Notice that if every experiment we ever did had a genuine effect then all positive tests would be correct and the false discovery rate would be zero.  Notice also that the tree diagram shows that 98% of negative tests give the right result: false negatives are rare. But that's not surprising because 90% of tests there really is no effect, so there's a good chance that a negative test will be right.  If there were no real effects at all, as in the homeopathic example, then 100% of negative tests would be right.

### Underpowered studies

The case just described is unusually good.  In practice many published results have a power far less than 0.8.  Values around 0.5 are common, and 0.2 is far from rare. Over half a century ago, Cohen (1962) found

*" . . . that the average power (probability of rejecting false null hypotheses) over the 70 research studies was 0.18 for small effects, 0.48 for medium effects, and 0.83 for large effects.*"

He was talking about social psychology.  He was largely ignored, despite the fact that many papers appeared in the statistical literature that discussed the problem.

Half a century later, Button *et al.* (2013) said

*"We optimistically estimate the median statistical power of studies in the neuroscience field to be between about 8% and about 31%"*

This is disastrously low.  It is no better than it was 50 years ago.  That's because many effects are quite small, and inadequate sample sizes are used, and because the warnings of statisticians have been ignored.

We can easily look at the consequences of low power either from tree diagrams as in Fig 2, or by simulating many *t* tests.  The examples of *t* tests shown in Figs 3 – 5 were simulated with 16 observations in each group. That was enough to give a power of 0.78, close to the value often used in the design of clinical trials.  If we use only 8 observations in each group, the power of the test falls to 0.46 and with 4 observations in each group, the power is only 0.22, so there is only a 22% chance of detecting a real effect when it's there.  Sadly values like these, though obviously unsatisfactory, are only too common.

The distribution of the 'observed' *P* values for a power of 0.22 is shown in Fig 6.  It's far more spread out than in Fig. 5b, with only 22% of *P* values equal to or less than 0.05, as would be expected from the power of the tests.  If a 'significant' test occurs, the next test will be 'not significant' in 78% of cases.  Again we see that *P* values are not at all reproducible. The fact that the *P* value will often differ greatly when the experiment is repeated is sometimes used as a criticism of the whole *P* value approach.  In fact it's to be expected and the conventional tests do exactly what it says on the tin.  This phenomenon is illustrated graphically in the *Dance of the P Values* (http://www.youtube.com/watch?v=ez4DgdurRPg )

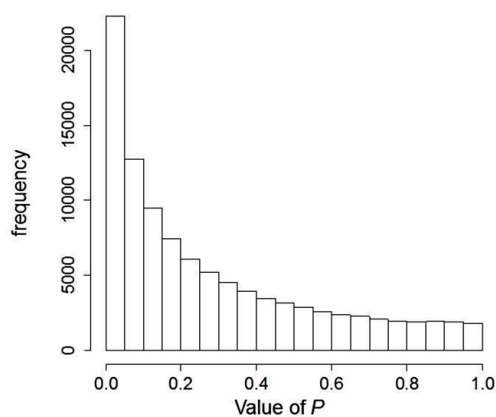

**Figure 6.**  Distribution of 100,000 *P* values from tests like those in Fig 5, but with only 4 observations in each group, rather than 16.  The calculated power of the tests is only 0.22 in this case, and it's found, as expected, that 22% of the *P* values are equal to or less than 0.05.

### *The inflation effect*

If you do manage to get a positive 'significant' difference ($P \leq 0.05$) with a power of 0.46, and you look at the size of the effect for those results for that are 'significant', it comes out as about 1.4. And with a power of 0.22 it comes out as about 1.8 (both are bigger than the true difference between means of 1.0). In other words, not only do you mostly fail to detect the true effect, but even when you do (correctly) spot it, you get its size wrong. The inflation effect gets really serious when the power is low. The estimated effect size is almost twice its true value with a power around 0.2. That's because the test is more likely to be positive in the small number of experiments that show a larger than average effect size. The effect inflation, as a function of power (or of the number of observations in each group), is plotted in Fig 7.

There is no simple way to calculate the size of the effect inflation, but the R script (Colquhoun, 2014b) allows you to estimate the effect inflation for numbers that are appropriate for your problem

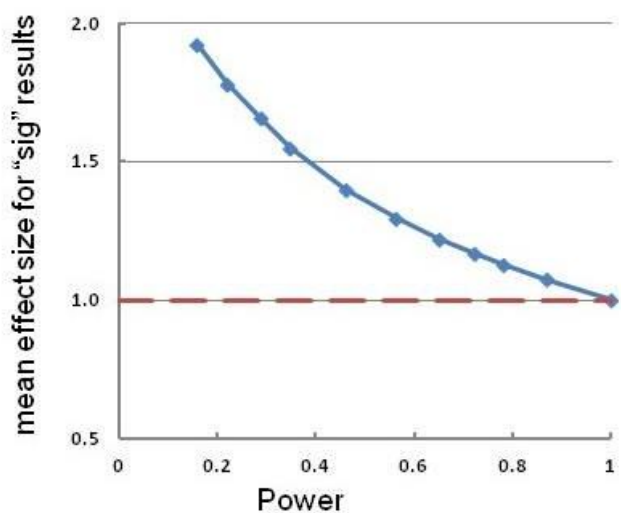

**Figure 7** The average difference between means for all tests that came out with $P \leq 0.05$. Each point was found from 100,000 simulated $t$ tests, with data as in Fig 4. The power of the tests was varied by changing the number, $n$, of 'observations' that were averaged for each mean. This varied from $n = 3$ (power = 0.157) for the leftmost point, to $n = 50$ (power = 0.9986) for the rightmost point. Intermediate points were calculated with $n = 4, 5, 6, 8, 10, 12, 14, 16$ and 20.

Most seriously of all though, your chance of making a fool of yourself increases enormously when experiments are underpowered. Even in the best case, where half your experiments have a true effect, you'll make a fool of yourself by claiming an effect is real when it's not in about 10% of cases for a power of 0.5 (sample size of $n = 8$ in the simulations), and in 20% of tests for a power of 0.2 ($n = 4$ in the simulations). If only 10% of your experiments have a true effect, as illustrated in Figure 2, you'll make a fool of yourself in almost 50% of cases when the power is about 0.5, and in a staggering 70% of cases when the power is only 0.2.

The spreadsheet with the results for simulated $t$ tests, and the R programme that allows you to run them yourself, are available (Colquhoun, 2014b).

### Two more approaches

It is already clear that if you use $P = 0.05$ as a magic cut off point, you are very likely to make a fool of yourself by claiming a real effect when there is none. This is particularly the case when experiments are underpowered. In every case we've looked at so far, the probability of incorrectly declaring an effect to be real has been greater than 5%. It's varied from 6% to a disastrous 70%, depending on the power of the test, and depending on the proportion of experiments we do over a lifetime in which there is a real effect: the smaller this proportion, the worse the problem.

It's easy to see why this happens. If many of the tests you do have no real effect then the large number of false positives they generate overwhelms the number of true positives that come from the small number of experiments where there is a genuine effect.

All the results so far have referred to conventional tests of significance in which you claim to have discovered a real effect when you observe $P \leq 0.05$ (or some other specified value). The results are already alarming. But there is another subtlety to be considered.

### *What happens if we consider P = 0.05, rather than P ≤ 0.05 ?*

One conventionally declares a result to be 'significant' if $P$ is equal to *or less than* 0.05 (or some other specified value). Thus $P = 0.047$ would be described as "significant" in the classical Fisherian method. Some statisticians would say that, once you have observed, say, $P = 0.047$, that is part of the data so we should not include the *or less than* bit. That is indisputable if we are trying to interpret the meaning of a single test that comes out with $P = 0.047$. To interpret this we need to see what happens in an imaginary series of experiments that all come out with $P$ near to 0.05.

It's easy to see what happens by using repeated simulations of $t$ tests, as above, but this time we restrict attention to only those tests that come out close to 0.05. We run the same simulations as before, but rather than looking at all experiments for which $P$ is 0.05 *or less*, we confine our attention to only those experiments that come out with a $P$ value between 0.045 and 0.05. Arguably, this is what we need to do in order to interpret a single experiment that produces $P = 0.047$.

When we run the simulations for tests with reasonable power (a sample size of $n = 16$, giving power near to 80%, as in Figs 2 – 5), when there is a real effect of the size illustrated in Fig 4, we find that out of 100,000 tests, 1424 come out with a $P$ value between 0.045 and 0.05 (true positives). And when we run the simulations again, with no real effects (the true mean difference between treatment and control and control is zero), we find that 511 tests come out with a $P$ value between 0.045 and 0.05 (false positives). So there are 1935 positive tests, of which 511 (26%) are false positives. This is the most optimistic case, in which the power is adequate and it's assumed that half your experiments have a true effect and half didn't.

Thus, if you observe a $P$ value close to 0.05 and declare that you've discovered a real effect, you'll make a fool of yourself 26% of the time, even in the most optimistic case.

Interestingly, this percentage doesn't change much when tests are underpowered (it's already a disastrously high false discovery rate).

If we look at the case where most (90%) of experiments have no real effect, as in Fig 2, the outcome gets even worse. If we look at only those experiments that give a $P$ value between 0.045 and 0.05 we find that in 76% of these 'just significant' experiments, there was in fact no real effect: the null hypothesis was true. Again, this number is almost independent of power.

The outcome is that if you declare that you've made a discovery when you observe a $P$ value close to 0.05, you have at the least a 26% chance of being wrong, and often a much bigger chance. Yet many results get published for which the false discovery rate is at least 30%. No wonder there is a problem of reproducibility.

These statements refer only to tests that come out close to 0.05, so they don't tell you about the number of times you make a fool of yourself over a lifetime (not all your results will be close to 0.05), but they do indicate that observation of $P$ close to 0.05 tells you remarkably little about whether or not you've made a discovery.

### Berger's approach

In order to do these calculations we've had to postulate the prevalence of tests that we do in which there is in fact a genuine effect (null hypothesis untrue). We have seen that even in the most optimistic case, in which prevalence is 0.5, the false discovery rate is alarmingly high. A Bayesian would refer to the prevalence as the prior probability that there is a real effect. There is no need to describe it in this way. It's a normal frequentist probability, which could, in principle, be estimated by sufficiently rigorous tests.

The problem with asking a Bayesian about what to do is that you may get as many different answers as there are Bayesians. James Berger devised an ingenious solution to this problem (Berger & Sellke,.1987; Sellke *et al.*, 1981). He gave a result that applies regardless of what the shape of the prior distribution might be. In effect, it chooses the prior distribution that is most favourable to the hypothesis that there is a real effect. Using this one can calculate the minimum false discovery rate that corresponds to any observed $P$ value. For $P = 0.05$ the false discovery rate calculated in this way is 0.289. This is a minimum value. It means that if you observe a $P$ value that is close to 0.05, there is at least a 29% chance that there is in fact no real effect. This result is quite close to the false discovery rates that were inferred from the simulated $t$ tests above, when we confined our attention to experiments that gave $P$ values between 0.045 and 0.05. More information is given in the Appendix.

If you believe that it's unacceptable to make a fool of yourself 30% of the time, what should you do? According to Berger's approach, a $P$ value close to 0.001 corresponds to a false discovery rate of 1.84 percent (see appendix). If you believe that it's tolerable to take a 1.8 percent risk of making a fool of yourself, then you don't claim to have discovered a real effect in an experiment that gives $P$ bigger than 0.001.

This procedure amounts roughly to adopting a 3-sigma policy, rather than a 2-sigma rule. Two standard deviations from the mean excludes about 5% of the area under a normal distribution (2.5% in each tail), and that's what happens when you use $P = 0.05$. Three standard deviations from the mean excludes 0.27% of the area. Berger's calibration suggests that $P = 0.0027$ corresponds to a false discovery rate of 0.042, not far from the 0.05 level that is customarily abused.

### Is the argument Bayesian?

The way that you predict the risk of getting a false positive is often described as being an application of Bayes' theorem. There is a fascinating argument among statisticians about the practical use of Bayes' theorem. The argument started after Bayes' results were published in 1764, and it rages on still. One of the controversial bits about using Bayesian methods is the necessity to abandon the easy definition of probability as a long run frequency, and instead to consider it as subjective betting odds. The other is the need to specify how strong your belief in the outcome is *before* the experiment is done (a prior probability), an exercise that can come dangerously close to feeding your prejudices into the result.

Luckily, though, it isn't necessary to get involved in any of these subtleties.

I maintain that the analysis here may bear a formal similarity to a Bayesian argument, but is free of the more contentious parts of the Bayesian approach. The arguments that I have used contain no subjective probabilities, and are an application of obvious rules of conditional probabilities.

The classical example of Bayesian argument is the assessment of the evidence of the hypothesis that the earth goes round the sun. The probability of this hypothesis being true, given some data, must be subjective since it's not possible to imagine a population of solar systems, some of which are heliocentric and some of which are not. The solar system is either heliocentric or not: it can't be 95% heliocentric.

One can similarly argue that an individual drug either works or it doesn't (disregarding some obvious assumptions that underlie that statement). But the need for subjective probabilities vanishes if we think of a lifetime spent testing a series of drugs, one at a time, to see whether or not their effects differ from a control group. It's easy to imagine a large number of candidate drugs some of which are active (fraction $P(real)$ say) , some of which aren't. So the prevalence (or prior, if you must) is a perfectly well-defined probability, which could be determined with sufficient effort. If you test one drug at random, the probability of it being active is $P(real)$. It's no different from the probability of picking a black ball from an urn that contains a fraction $P(real)$ of black balls. to use the statisticians' favourite example.

That way of looking at the problem is exactly analogous with the case of screening tests, which certainly does not necessitate subjective probabilities

### Conclusions: what can be done?

All of the approaches above suggest that if you use $P = 0.05$ as a criterion for claiming that you have discovered an effect you'll make a fool of yourself at least 30% of the time. This alone implies that many published claims are not true.

It's important to notice that the calculations described here are the most optimistic view possible. They apply to properly designed tests in which treatments are randomly allocated to groups, there is no bias (e.g. assessments are blinded) and all negative results are published. It is also assumed that there is a single pre-specified outcome, so there is no problem arising from multiple comparisons. In real life, such perfect experiments are rare. It follows that 30% is very much a minimum for the proportion of published experiments which wrongly claim to have discovered an effect. To that extent, Ioannidis' assertion that "most published research findings are false" seems to be not unduly alarmist.

The blame for the crisis in reproducibility has several sources.

One of them is the self-imposed publish-or-perish culture (Colquhoun, 2011), which values quantity over quality, and which has done enormous harm to science.

The mis-assessment of individuals by silly bibliometric methods has contributed to this harm. Of all the proposed methods, 'altmetrics' is demonstrably the most idiotic Colquhoun & Plested, (2014a), Yet some vice-chancellors have failed to understand that (Colquhoun, 2013b)

Another cause of problems is scientists' own vanity, which leads to the PR department issuing disgracefully hyped up press releases. (Colquhoun, 2013c)

In some cases, the abstract of a paper even states that a discovery has been made when the data say the opposite. This sort of spin is common in the quack world. Yet referees and editors get taken in by the ruse (e.g. see a study of acupuncture: Colquhoun, 2013a).

The reluctance of many journals (and many authors) to publish negative results biases the whole literature in favour of positive results. This is so disastrous in clinical work that a pressure group has been started; altrials.net "All Trials Registered: All Results Reported".

Yet another problem is that it has become very hard to get grants without putting your name on publications to which you have made little contribution. This leads to exploitation of young scientists by older ones (who fail to set a good example). It has led to a slave culture in which armies of post doctoral assistants are pushed into producing more and more papers for the glory of the boss and the university, so they don't have time to learn the basics of their subject (including statistics). Peter Lawrence (2007) has set out the problems in *The Mismeasurement of Science*.

And, most pertinent to this paper, a widespread failure to understand properly what a significance test means must contribute to the problem.

Here are some things that can be done.

- Notice that all statistical tests of significance assume that the treatments have been allocated at random. This means that application of significance tests to observational data, e.g. epidemiological surveys of diet and health, is not valid. You can't expect to get the right answer. The easiest way to understand this assumption is to think about randomisation tests (which should have replaced *t* tests decades ago, but which are still rarely used). There is a simple introduction in *Lectures on Biostatistics* (Colquhoun. 1971, chapters 8 and 9). There are other assumptions too, about the distribution of observations, independence of measurements), but randomisation is the most important.

- Never, *ever*, use the word "significant" in a paper. It is arbitrary, and, as we have seen, deeply misleading. Still less should you use "almost significant", "tendency to significant" or any of the hundreds of similar circumlocutions listed by Matthew Hankins (2013) on his *Still not Significant* blog.

- If you do a significance test, just state the *P* value and give the effect size and confidence intervals. But be aware that 95% intervals may be misleadingly narrow, and they tell you nothing whatsoever about the false discovery rate. Confidence intervals are just a better way of presenting the same information that you get from a *P* value.

- Observation of a *P* value close to 0.05 means nothing more than 'worth another look'. In practice, one's attitude will depend on weighing the losses that ensue if you miss a real effect against the loss to your reputation if you claim falsely to have made a discovery.

- Do some rough calculations of the sample size that might be needed to show a worthwhile effect, Underpowered studies still abound and contribute to both high false discovery rates and effect-size inflation.

- If you want to avoid making a fool of yourself too often, don't regard anything bigger than *P* < 0.001 as a demonstration that you've discovered something. Or, slightly less stringently, use a three-sigma rule.

Similar conclusions have been reached, for similar reasons, by many others, e.g. Sterne & Davey Smith (2001) and Valen Johnson (2013). But they have been largely ignored by authors and editors. One exception to that is genome-wide association studies, which were notorious for false positive associations in the early days, but which have now learned the statistical lesson (e.g. Bush & Moore, 2012) Nevertheless, the practice of labelling a difference between two values with an asterisk, and saying it's a discovery is still rampant in the biomedical literature. No wonder so much of it is wrong.

One must admit, however reluctantly, that despite the huge contributions that Ronald Fisher made to statistics, there is an element of truth in the conclusion of a perspicacious journalist

*"The plain fact is that 70 years ago Ronald Fisher gave scientists a mathematical machine for turning baloney into breakthroughs, and °flukes into funding. It is time to pull the plug".* Robert Matthews Sunday Telegraph, 13 September 1998.

## APPENDIX

All the calculations here are based on the rules of conditional probability (a simple introduction is given by Colquhoun, 1971, section 2.4). The probability of observing both event A and event B (left hand term) can be written in two ways.

$$P(A \cap B) = P(A|B)P(B) = P(B|A)P(A) \qquad \text{(A1)}$$

$P$(A|B) is read as the probability of A given that B has occurred. If A and B are independent, the probability of A occurring is the same whether or not B has occurred so $P$(A|B) is simply $P$(A) and eqn. A1 reduces to the multiplication rule of probability. It follows from eq. A1 that

$$P(A|B) = \frac{P(A)P(B|A)}{P(B)}$$

$$\text{(A2)}$$

## The screening example

Define event A to mean you have the condition which is being screened for, so A = *ill*. Define event B as meaning that your diagnostic test comes out positive. The probability that you are actually ill given that the test comes out positive, i.e.

$$P(ill|test\ pos) = \frac{P(ill)P(test\ pos|ill)}{P(test\ pos)}$$

The false discovery rate is therefore

$$P(not\ ill|test\ pos) = 1 - \frac{P(ill)P(test\ pos|ill)}{P(test\ pos)}$$

The probability that you test positive, $P$(B), can be expressed thus

$$P(B) = P(B|A)P(A) + P(B|notA)P(notA)$$

So the fraction of "significant" tests in which you really are ill is

$$P(ill|test\ pos)$$
$$= \frac{P(ill)P(test\ pos|ill)}{P(test\ pos|ill)P(ill) + P(test\ pos|not\ ill)P(not\ ill)}$$

$$\text{(A3)}$$

And the false discovery rate is

$$P(not\ ill|test\ pos)$$
$$= \frac{P(test\ pos|not\ ill)P(not\ ill)}{P(test\ pos|ill)P(ill) + P(test\ pos|not\ ill)P(not\ ill)}$$

This expresses, as an equation, exactly what we inferred from the tree diagram in Fig 1. For example, in Fig 1 we postulated that the prevalence of MCI in the population is 1%, so

$$P(ill) = 0.01$$

$$P(not\ ill) = 0.99$$

The sensitivity of the test was 80%, so 80% of people who have MCI will test positive

$$P(test\ pos|ill) = 0.8$$

The specificity of the test was 0.95. *i.e.* 95 percent of people without the condition will be correctly diagnosed as not having it

$$P(test\ neg|not\ ill) = 0.95$$

So

$$P(test\ pos|not\ ill) = 1 - 0.95 = 0.05$$

Thus

$$P(ill|test\ pos) = \frac{0.01 \times 0.8}{0.01 \times 0.8 + 0.05 \times 0.99} = 0.139$$

Thus the false discovery rate of the test is 1 - 0.139 = 86.1% false positives, as found from the tree diagram in Fig 1.

## Significance tests.

The argument is much the same as for screening. Denote as *real* the event that there is a real difference between test and control, i.e. the null hypothesis is false. This is the same as the prevalence, in the screening test calculations. So *not real* means the null hypothesis is true –there is no real effect. Denote as *test sig* the event that the test indicates the result is 'significant', i.e. comes out with $P \le 0.05$ (or whatever level is specified). We can now work through the example in Fig 2 with equations. From A2

$$P(real|test\ sig)$$
$$= \frac{P(real)P(test\ sig|real)}{P(test\ sig|real)P(real) + P(test\ sig|not\ real)P(not\ real)}$$

$$\text{(A4)}$$

As in Fig 2, suppose that, in a series of tests, 10% have real effects but in 90% the null hypothesis is true. Thus

$$P(real) = 0.1$$

$$P(not\ real) = 0.9$$

Say the power of the test is 0.8, so when there is a real effect it will be detected 80% of the time.

$$P(test\ sig|real) = power = 0.8$$

And the conventional 'significance' level is 0.05, so

$$P(test\ sig|not\ real) = 1 - 0.95 = 0.05$$

Putting these values into eq. A4

$$P(real|test\ sig) = \frac{0.1 \times 0.8}{0.1 \times 0.8 + 0.05 \times 0.90} = 0.64$$

This is the same result that we got from the tree diagram in Fig 2. Even for this well-powered test, the null hypothesis is true in $1 - 0.64 = 36\%$ of tests that are declared 'significant'. That sort of false discovery rate means that you would make a fool of yourself in a bit more than 1 in 3 cases in which you claim to have discovered an effect.

### Bayes' factor –posterior odds

Another approach to deciding whether or not an effect is real is to look at the *likelihood ratio*. This term likelihood, in its statistical sense, means probability of observing the data, given a hypothesis, *I.e.* the probability of observing the data if that hypothesis were true. Say the two hypotheses to be compared are $H_0$ and $H_1$ where $H_0$ means that there is no real difference between treatment and control (the null hypothesis) and $H_1$ means that there is a real difference. The likelihood ratio is

$$B = \frac{P(test\ sig|H_0)}{P(test\ sig|H_1)}$$

(this is also known as the Bayes' factor, but there is no need to consider it as a Bayesian concept). The bigger this is, the more the data favours the $H_1$, the greater the likelihood that a real difference exists.

It follows from A2 that the false discovery rate is

$$P(not\ real|test\ sig)$$
$$= \frac{P(test\ sig|not\ real)P(not\ real)}{P(test\ sig|real)P(real) + P(test\ sig|not\ real)P(not\ real)}$$

Thus the odds ratio for $H_0$ (*versus* $H_1$) is the ratio of these two quantities, i.e.

$$odds\ ratio = \frac{P(not\ real|test\ sig)}{P(real|test\ sig\ )}$$
$$= \frac{P(H_0)}{P(H_1\ )} \frac{P(test\ sig|H_0)}{P(test\ sig|H_1\ )} = \frac{P(H_0)}{P(H_1)} B$$

Thus the odds on $H_0$ are given by the product of the prior odds, $P(H_0)/P(H_1)$ and the likelihood ratio, $B$.

### Numerical example

We'll use, once again, the numbers that were used in Fig. 2. The prior odds ratio is

$$\frac{P(H_0)}{P(H_1)} = \frac{9}{1}$$

i.e. there is a 9 to one chance that the null hypothesis is true, so in $9/(9+1) = 90\%$ of tests the null hypothesis is true, and in the remaining 10% there is a real effect to be discovered. As before, using $P = 0.05$ as cut off

$$P(test\ sig|H_0) = 0.05$$

And the power of the test is 0.8

$$P(test\ sig|H_1) = 0.8$$

The likelihood ratio for $H_0$ versus $H_0$

$$L = \frac{0.05}{0.8} = \frac{1}{16}$$

Thus

$$OR = odds\ ratio\ on\ H_0\ vs\ H_1 = \frac{9}{1} \frac{0.05}{0.8} = 0.5625$$

So the odds that there is really an effect are less than 2 to 1. Put another way, the false discovery rate is

$$P(H_0|test\ sig) = P(not\ real|test\ sig) = \frac{OR}{1 + OR}$$
$$= \frac{0.5625}{1 + 0.5625} = 0.36$$

Once again, we find a false discovery rate of 36%, far bigger than the $P = 0.05$ used for the test.

### The Berger approach

James Berger and colleagues proposed to solve the problem of the unknown prior distribution by looking for a lower bound for the likelihood ratio (Bayes factor), for $H_0$ relative to $H_1$. Expressed as a function of the observed $P$ value, Sellke *et al*, (2001) suggest

$$B(p) = -eP\ \log(P)$$

(this holds for $P < 1/e$, where $e = 2.71828 . . .$). This is the smallest odds against the null hypothesis $H_0$ that can be generated by any prior distribution, whatever its shape. Therefore it is the choice that most favours the rejection of the null hypothesis. The odds can be converted to an equivalent probability

$$\alpha(P) = \frac{B(P)}{1 + B(P)} = \frac{1}{1 + \frac{1}{(-eP\log(P))}}$$

(A5)

For the case where the prior probability of having a real effect is $P(H_1) = P(H_0) = 0.5$, this can be interpreted as the minimum false discovery rate (Sellke *et al.*, 2001). It gives the *minimum* probability that, when a test is 'significant', the null hypothesis is true: i.e. it is an estimate of the minimum false discovery rate, or false positive rate. Berger refers to it as the conditional error probability. Some values are given in Table A1

| $P$ | 0.2 | 0.1 | 0.05 | 0.01 | 0.005 | 0.001 |
|---|---|---|---|---|---|---|
| $\alpha(P)$ | 0.465 | 0.385 | 0.289 | 0.111 | 0.067 | 0.0184 |

**Table A1** *P* values and their corresponding conditional error probabilities, $\alpha(P)$, calculated from equation A5, as in (Sellke *et al.* 2001).

The minimum false discovery rate for $P = 0.05$ is seen to be 0.289. In other words, if you claim you have discovered something when you observe a *P* value close to 0.05, you will make a fool of yourself in about 30% of cases.

In the example used in Fig 2 the false discovery rate was 36%, which is compatible with Berger's result, but this isn't strictly comparable, because Fig 2, and the first set of simulations looked at all tests which came out with $P \leq 0.05$. Berger's approach concerns only those tests that come out close to the specified value, $P = 0.05$ in the example. In the second set of simulations we looked at tests that gave *P* values between 0.045 and 0.05. These gave a false discovery rate of *at least* 26% (in the case where the prior probability of a real effect was 0.5) and a false discovery rate of 76% in the case, as in Fig 2, when only 10% of the experiments have a real effect. These results are close to Berger's assertion that the false discovery rate will be at least 29% regardless of what the prior distribution might be.

### Acknowledgments

I'm very grateful to the following people for their comments on earlier drafts of this paper.

Prof Stephen Senn (Competence Center in Methodology and Statistics, CRP-Santé, Luxembourg), Drs Ioanna Manolopoulou and Simon Byrne (Statistics, UCL), Dr Harvey Motulsky (CEO, Graphpad Software Inc), Prof Dorothy Bishop (Oxford) and Prof. Lucia Sivilotti (UCL). If, despite their best efforts, I have made a fool of myself, blame me not them.

I'm grateful to both referees for their helpful comments. I'm particularly grateful to referee 2 for allowing that after having "constructed such a sturdy soapbox it would be a shame not to let the author speak from it"